# Comparing the Impact of Pedagogy-Informed Custom and General-Purpose GAI Chatbots on Students' Science Problem-Solving Processes and Performance Using Heterogeneous Interaction Network Analysis

Hanyu Su[1[0009-0002-2947-6056]], Huilin Zhang[1[0009-0008-0120-3017]] and Shihui Feng[1[0000-0002-5572-276X]]

[1] Faculty of Education, The University of Hong Kong, Hong Kong 999077, China
{suhanyu, huilinzhang}@connect.hku.hk, shihuife@hku.hk

**Abstract.** Problem solving plays an essential role in science education, and generative AI (GAI) chatbots have emerged as a promising tool for supporting students' science problem solving. However, general-purpose chatbots (e.g., ChatGPT), which often provide direct, ready-made answers, may lead to students' cognitive offloading. Prior research has rarely focused on custom chatbots for facilitating students' science problem solving, nor has it examined how they differently influence problem-solving processes and performance compared to general-purpose chatbots. To address this gap, we developed a pedagogy-informed custom GAI chatbot grounded in the Socratic questioning method, which supports students by prompting them with guiding questions. This study employed a within-subjects counterbalanced design in which 48 secondary school students used both custom and general-purpose chatbot to complete two science problem-solving tasks. 3297 student-chatbot dialogues were collected and analyzed using Heterogeneous Interaction Network Analysis (HINA). The results showed that: (1) students demonstrated significantly higher interaction intensity and cognitive interaction diversity when using custom chatbot than using general-purpose chatbot; (2) students were more likely to follow custom chatbot's guidance to think and reflect, whereas they tended to request general-purpose chatbot to execute specific commands; and (3) no statistically significant difference was observed in students' problem-solving performance evaluated by solution quality between two chatbot conditions. This study provides novel theoretical insights and empirical evidence that custom chatbots are less likely to induce cognitive offloading and instead foster greater cognitive engagement compared to general-purpose chatbots. This study also offers insights into the design and integration of GAI chatbots in science education.

**Keywords:** Custom GAI Chatbot, Science Problem Solving, Heterogeneous Interaction Network Analysis, Human-AI Interaction

# 1 Introduction

Problem solving is a critical 21st-century skill [1] and an important instructional tool in science education [2]. Prior studies have shown the potential of chatbots to support students' science problem solving by engaging learners in natural language dialogues [3, 4]. However, earlier AI chatbots were largely rule-based or developed on intent-based platforms (e.g., Dialogflow), which constrained their adaptability and flexibility [5]. Recent advances in generative AI (GAI) have substantially improved the conversational competence and knowledge capacity of chatbots [6], enabling more flexible human-AI interactions and adaptive support for students' science problem solving. Nevertheless, research in this area remains at an early exploratory stage [4].

Since the emergence of GAI, most studies have used general-purpose GAI chatbots (e.g., ChatGPT, DeepSeek) to support science learning, whereas limited studies have examined custom GAI chatbots [4, 5]. A key limitation of general-purpose chatbots is that they are not primarily oriented toward educational goals [8]; instead, they are optimized to execute users' commands. Prior research has suggested that such design may provide students with cognitive shortcuts [7], potentially fostering over-reliance on AI-generated answers. To address this concern, we designed a custom chatbot grounded in the Socratic questioning method [9]. Rather than delivering ready-made answers, it guides students through iterative questioning to promote meaningful interaction and cognitive engagement throughout the problem-solving processes. However, how pedagogy-informed custom chatbots, compared with general-purpose chatbots, differentially influence students' science problem-solving processes and performance remains underexplored. Therefore, this study aims to address this gap by comparing these two types of GAI chatbots with respect to students' problem-solving processes and performance. Accordingly, this study is guided by the following three research questions:

RQ1: How do students' interaction intensity and cognitive interaction diversity differ between using the pedagogy-informed custom and general-purpose GAI chatbots during science problem solving?

RQ2: How do students' problem-solving patterns differ between using the pedagogy-informed custom and general-purpose GAI chatbots during science problem solving?

RQ3: How does students' problem-solving performance differ between using the pedagogy-informed custom and general-purpose GAI chatbots?

This study provides novel theoretical insights and empirical evidence for understanding Human-AI Interactions (HAI) in science problem-solving contexts and sheds light on the design of GAI-powered chatbots to support science problem solving.

# 2 Literature Review

## 2.1 Problem Solving in Science Education

Problem-solving is a process that includes utilizing previously learned knowledge and understanding to fulfill the requirements of unfamiliar scenarios [10]. It is pivotal in science education, equipping students with the ability to handle complex situations and fostering inquiry skills essential for future scientists [2]. Problem-solving performance is often measured by task products like solution outcomes or scientific papers [11, 12]. Furthermore, problem solving is characterized as a sequence of cognitive processes that entails intellectual calculations [1]. This process has been conceptualized through various frameworks, from Pólya's problem solving techniques to Ervynck's three-stage progression [13, 14]. A comprehensive understanding of scientific problem-solving process empowers educators to disaggregate this intricate phenomenon into discrete, manageable components. This analytical approach facilitates the precise identification of student learning outcomes and areas of deficiency, thereby enabling the provision of targeted and timely pedagogical interventions.

Previous studies have employed various types of data to understand students' problem-solving processes. For instance, Tschisgale et al. utilized process mining and sequence analysis to investigate physics problem-solving, requiring students to verbalize strategies and concepts, which were then coded into categories like assumptions and conceptual aspects [12]. Some studies, such as [15] and [16], have investigated the strategies employed by students in solving chemistry problems through retrospective think-aloud protocols. Event log data from digital learning environments have also been leveraged to analyze science problem-solving processes. The Trends in International Mathematics and Science Study (TIMSS) exemplifies research on science problem solving by employing computer-based simulations of real-world contexts. Through interactive mechanisms like data entry and experimental manipulation, the system assesses fundamental inquiry skills ranging from hypothesis generation to evidence-based argumentation [17]. Similarly, Scherer et al. documented students' click and entry operations in identifying unknown chemicals within a virtual environment, assessing the students' complex problem-solving process across four dimensions [18].

### 2.2 AI Chatbots in Science Education

AI chatbots facilitate personalized, round-the-clock interaction with students via natural language, embodying diverse educational personas such as teaching, peer, teachable, and motivational agents [19]. Research on chatbots has largely focused on their various applications (e.g., facilitating practice, disseminating knowledge, supervising learning activities, and offering emotional scaffolding [20]), the technological underpinnings of their design (e.g., rule-based systems, statistical models, machine learning, large language models [21]), and their inherent potential and associated challenges.

In science education, the abstract nature of scientific concepts and the high cognitive load of complex problem-solving necessitate students' access to timely support. Consequently, AI chatbots have emerged as particularly prominent tools, specifically designed to address these challenges by rectifying conceptual misconceptions through dialogue, providing immediate feedback on complex quantitative problems, decomposing abstract principles, and liberating teacher capacity for higher-order instruction [4]. More recently, with the advent of GAI's superior text generation quality and enhanced flexibility, researchers are exploring its capabilities to provide students with personalized scientific materials, explain scientific concepts, create graphical representations, and offer tailored suggestions for improvement [21, 22].

AI chatbots can potentially impact students' scientific learning, encompassing both outcomes (e.g., scientific performance, learning satisfaction [23]) and processes (e.g., learning engagement [24]). However, the assessment of chatbots' effectiveness on these learning aspects relies predominantly on students' subjective reflections (self-report survey or interviews) and on tests; evaluations focusing on their learning behaviors or psychological factors remain scarce. Calvo-Utrilla et al. also emphasize that empirical data beyond subjective opinions are requisite to substantiate the effectiveness of chatbots [4]. Few studies began to explore how human and AI interact with each other. For example, Min et al. conducted a statistical analysis of student-GAI dialogue content and selected a subset of students for a case study to understand how different interaction patterns affect students' scientific inquiry abilities [25].

While AI offers multifaceted support in science education, it risks diminishing student cognitive agency by supplanting rather than scaffolding the cognitive subject [26]. To ensure AI fosters cognitive development, emerging research explores customized AI applications in science education. For instance, Tang and Putra developed a GAI chatbot framed by Bakhtin's theory of heteroglossia. By using prompts that positioned the AI as a dialogic partner, their study with 21 students showed the customized AI successfully elicited critical reflection, reasoning, and argumentation during interactions on socioscientific issues [27]. Similarly, Ng et al. compared GAI with rule-based chatbots. With tailored prompts, the GAI chatbot was more effective in improving students' scientific knowledge, behavioral engagement, and motivation [28].

Despite extensive inquiry into chatbot applications and GAI integration in science education, literature addressing how these agents scaffold higher-order thinking skills, particularly complex scientific problem-solving, remains limited. Given that successful problem-solving demands mobilizing domain knowledge and applying diverse cognitive skills [29], it is significant to explore how GAI can serve as an effective cognitive extension to assist students, rather than diminish their agency.

# 3 Methods

## 3.1 Participants

A total of 48 secondary school students from Hong Kong, China, participated in this study. They were Form 1 students aged between 11 and 14 years ($M$ = 12.06, $SD$ = 0.51). Participants were from two classes, with 18 students in Class A and 30 students in Class B. Ethical approval was obtained from the university's Human Research Ethics Committee. All participating students provided informed consent prior to the study.

### 3.2 Experimental Design

To control for potential confounding variables related to differences in students' prior abilities and learning characteristics, this study employed a within-subjects counterbalanced experimental design. Students were assigned to two groups based on their intact classes and completed two science problem-solving tasks using two different types of chatbots in a counterbalanced order. Specifically, as shown in Fig. 1, students in Class A first completed Task 1 using the general-purpose chatbot and then completed Task 2 using the custom chatbot, whereas students in Class B completed Task 1 using the custom chatbot followed by Task 2 using the general-purpose chatbot. All participating students were provided with an introduction to how to use the chatbot prior to beginning the science problem-solving tasks.

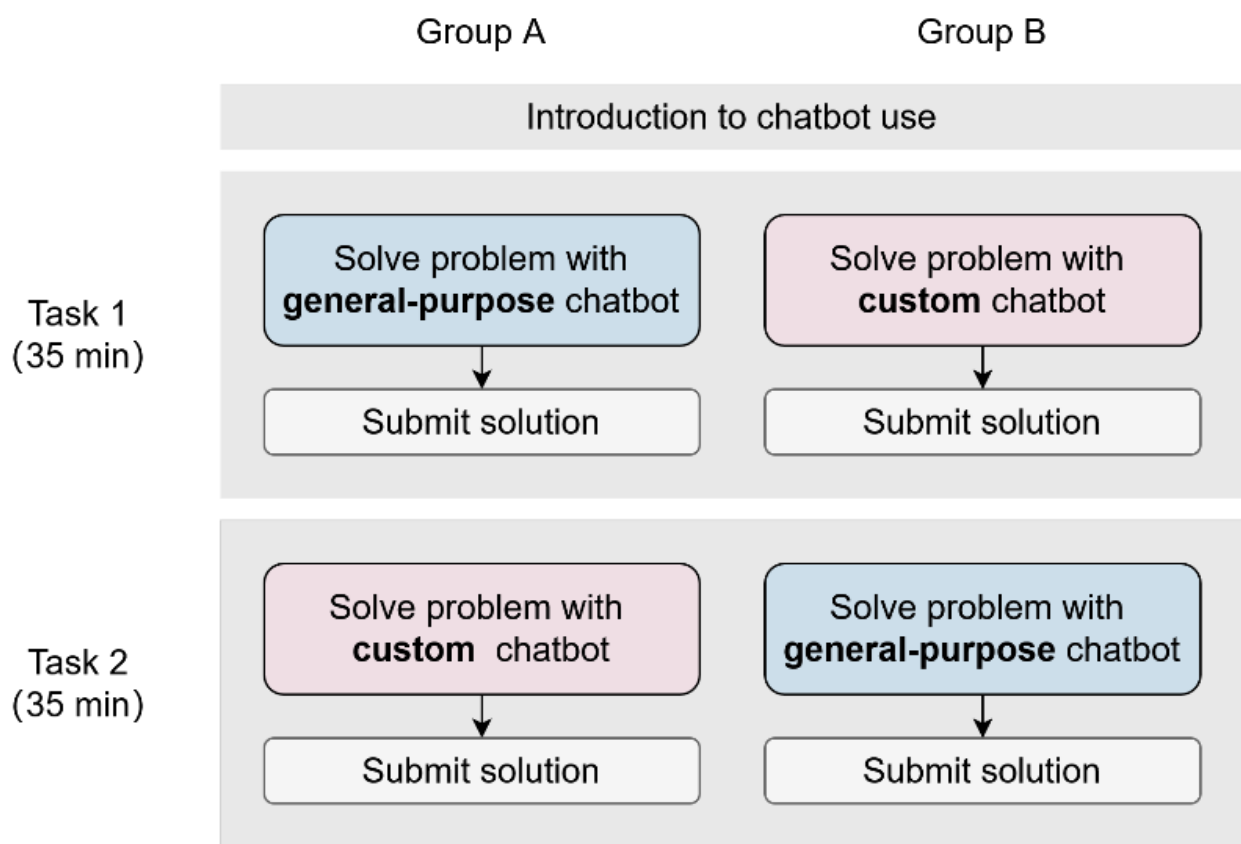

**Fig. 1.** The within-subject experiment design procedure.

Each problem-solving task session lasted 35 minutes. In Task 1 and Task 2, students were required to first learn basic concepts of measuring length or volume, derived from science curriculum textbooks, which they had not systematically studied before. Subsequently, they applied this newly gained knowledge to solve a related real-world problem, such as determining the dimensions of a box to fit a specific arrangement of ping-pong balls, or determining whether the water in a tank will overflow after adding a diamond. They were allowed to interact freely with the chatbot to solve the problems, but were not permitted to use any external resources beyond the given chatbot, such as textbooks or search engines. They were required to submit a solution to the given problem at the end of each task.

### 3.3 The Pedagogy-Informed Custom Chatbot and General-Purpose Chatbot

**Pedagogy-Informed Custom Chatbot Design.** Drawing on the Socratic questioning method [9], we designed and developed a custom chatbot. Socratic questioning emphasizes guiding students' thinking, reasoning, and discovery through systematic questioning rather than presenting direct answers. Consistent with this principle, our custom chatbot avoids giving answers directly; instead, it scaffolds students' problem-solving processes by delivering heuristic prompts that encourage active exploration. Through responding to these questions, students can engage in the learning process and understand concepts deeply. The interaction is iterative rather than single-turn. In addition to responding to questions, the chatbot generates follow-up questions, progressively scaffolding students' reasoning until they ultimately arrive at the final answer or solution.

We configured the large language model (LLM) using prompt engineering and Retrieval-Augmented Generation (RAG) techniques. Gemini 2.5 Flash was selected as the base model because, at the time, it offered strong performance and was cost-efficient. We carefully designed prompts to define the chatbot's response behaviors, as detailed in https://osf.io/wdmk4/overview?view_only=a991552056b04204937cf873de9207b1.

In addition, we employed RAG to familiarize the chatbot with our learning context, enabling it to generate responses that match learners' real learning materials and knowledge levels. We used students' science textbooks as the primary retrieval source. This approach ensured that the chatbot's responses were accurate, contextually relevant, and pedagogically appropriate.

The participants are non-native English-speaking students, despite their formal teaching language being English. Accordingly, the custom chatbot was prompted to always respond in English while providing a one-time translation. This feature can help reduce language-related barriers for students.

The chatbot was developed using React framework, offering a user-friendly frontend interface. As shown in Fig. 2, it adopts a classic chat-style interface that aligns with contemporary learners' interaction habits, avoiding redundant or unfamiliar operations.

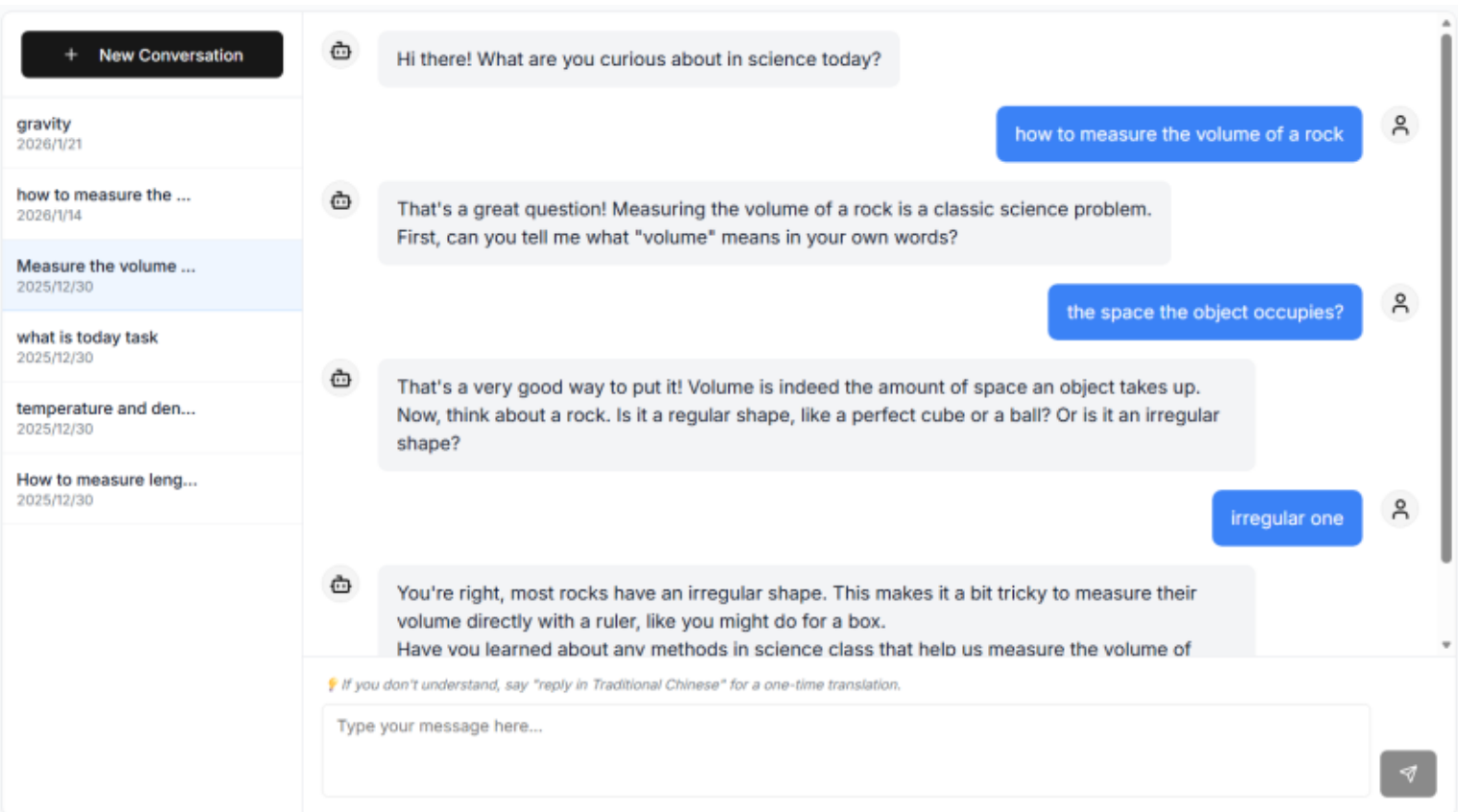

**Fig. 2.** The interface of the chatbot.

**General-Purpose Chatbot Design.** The general-purpose chatbot does not apply any prompt customization. To control for potential differences due to LLM base models, it employed the same Gemini 2.5 Flash model as the custom chatbot. Its interface was designed identically to that of the custom chatbot (see Fig. 2.) to eliminate confounding effects of device or interaction differences on comparative results.

### 3.4 Data Collection and Data Analysis

The system automatically recorded all student-chatbot dialogues to analyze problem-solving processes. In total, 3297 student-chatbot dialogues were collected. We also collected students' submitted solutions to examine their problem-solving performance.

**Coding Process.** In this study, to uncover students' problem-solving processes with the chatbot, we coded students' problem-solving interaction behaviors within the conversations. Prior research related to interaction process analysis has shown that when students engage in tasks with others or interactive agents, their interactions typically involve cognitive interactions and socio-emotional interactions [30, 31, 32]. We developed the coding scheme based on these two types of interactions. With the advent of GAI, students are now able to obtain direct answers from GAI with minimal effort. Therefore, to capture different levels of student engagement when interacting with chatbots, we further differentiated cognitive interactions into two subcategories: *Direct Requests* and *Exploratory Inquiry*. *Direct Requests* refer to interactions in which students explicitly ask the chatbot for answers or solutions, reflecting a relatively low level of cognitive engagement. In contrast, *Exploratory Inquiry* involves students actively engaging in reasoning, explanation seeking, fact recalling, or step-by-step problem solving with the chatbot, in which students demonstrate meaningful cognitive engagement. Additionally, *Socio-Emotional Interactions* include students' emotional expressions and socially oriented dialogue with the chatbot that serve interpersonal or affective functions. *Off-topic Interactions* were included to account for conversations that were not related to the focal science problem-solving tasks. Building on the specific problem-solving contexts in our experiment, we further identified concrete codes to operationalize each interaction category, as presented in Table 1.

**Table 1.** Coding scheme for analyzing students' problem-solving process interactions.

| Category | Code | Definition |
|---|---|---|
| Direct Request | Copy & Paste | Copy and paste the learning task to the chatbot directly |
| | Request direct answers/solutions | Request the chatbot to provide answers/solutions directly |
| | Request language translation | Request the chatbot to provide a translation |
| | Request formatting responses | Request the chatbot to adjust the response format |
| Exploratory Inquiry | Seek explanation of concept | Ask the chatbot questions to understand basic conceptual knowledge |
| | Explore solutions | Ask the chatbot questions to seek information to solve the problem |
| | Recall facts | Answer follow-up questions to recall factual knowledge |
| | Follow steps | Answer follow-up questions to understand concepts or solve problems |
| | Build-on questions | Ask questions based on the chatbot's response to extend ideas or deepen understanding |
| | Clarify information | Provide extra information or an explanation to clarify |
| | Evaluate | Summarize the solution and ask the chatbot to evaluate its correctness |
| | Refine solution | Adjust the solution according to the chatbot's feedback |
| Socio-Emotional Interactions | Self-disclosure | Share personal situations with the chatbot |
| | Emotional expression | Show emotions about the chatbot or the task |
| | Social engagement | Express appreciation, greetings, or politeness to keep conversation going |
| Off-topic Interactions | Irrelevant conversations | Talk about topics unrelated to the learning task |

Based on the initial coding scheme, two researchers independently coded 20% of the data. Interrater reliability was assessed using Cohen's kappa [33], which reached 0.918. Disagreements were resolved through discussion, and clarifications were added to the coding scheme. The remaining data were subsequently coded by one researcher.

**Student Problem-solving Heterogeneous Interaction Network Analysis (HINA).** HINA is a novel learning analytics framework that can yield multi-level insights into student-AI interactions, enabling a comprehensive understanding of students' meaningful interaction processes [34]. A heterogeneous interaction network (HIN) is a weighted bipartite graph consisting of multiple types of nodes, with edges connecting nodes from different sets [34]. It can be used to model complex interaction processes across diverse entities. In our context, HINA was employed to analyze the features and patterns of students' science problem-solving interactions in two conditions. Specifically, in this study we constructed a HIN for each condition: a student-interaction HIN $G = (V_s, V_c, E, \omega)$, where $V_s$ denoted the set of student nodes and $V_c$ represented the set of coded problem-solving interactions (the codes presented in Table 1). $E \subseteq V_s \times V_c$ represented the edges indicating which students exhibited which problem-solving interactions. The edge weight $w_{ij} = \omega((i,j))$ was defined as the frequency with which student $i$ engaged in problem-solving interaction $j$. The constructed HIN in each condition provided a clear and structured view of how students engaged in the problem-solving process in the custom and general-purpose GAI chatbots conditions. Based on the constructed HINs, this study employed the node-level analysis and dyadic-level analysis within the HINA analytical framework [34] to quantify the characteristics of students' problem-solving interactions as well as identify the significant associations within each condition.

**Analysis of Interaction Intensity and Cognitive Interaction Diversity (RQ1).** To address RQ1, we examined interaction features at the individual student level when using the custom and general-purpose chatbot. Two metrics from HINA's individual-level analysis [34] were used to measure students' interaction intensity and cognitive interaction diversity. Interaction intensity measures the extent to which a student engaged with the chatbot during science problem solving in each condition, operationalized as the total frequency of problem-solving interactions. It was quantified using the quantity measure proposed in HINA. Specifically, the interaction intensity of student $i$, denoted as $\mathrm{I}_i$, was defined as

$$\mathrm{I}_i = \sum_{j \in V_c} w_{ij} \tag{1}$$

The cognitive interaction diversity captures the extent to which students employed a variety of different cognitive interaction types rather than repeatedly relying on a single interaction strategy. Based on the student-interaction HIN described above, we constructed a student-cognitive interaction HIN subgraph $G' = (V_s, V_{c'}, E', \omega')$, where $V_{c'} \subseteq V_c$ includes nodes belonging to the *Direct Request* and *Cognitive Interactions* categories. This construct was quantified using the diversity measure proposed in HINA. Specifically, the cognitive interaction diversity of student $i$, denoted as $\mathrm{D}_i$, was defined as

$$\mathrm{D}_i = -\frac{1}{\log|V_{c'}|}\sum_{j\in V_{c'}}\left(\frac{w_{ij}}{s'_i}\right)\log\left(\frac{w_{ij}}{s'_i}\right) \quad (2)$$

where $s'_i = \sum_{j\in V_{c'}} w_{ij}$. The value of $\mathrm{D}_i$ ranges from 0 to 1, with higher values indicating a more balanced distribution across different types of cognitive interactions, suggesting a richer repertoire of cognitive strategies. The calculation for the two measures was conducted through the HINA Web Tool (https://hina-network.com) [35].

For each measure, the normality of the paired differences between the two chatbot conditions was assessed. Interaction intensity, whose differences violated the normality assumption, was analyzed using the non-parametric Wilcoxon signed-rank test, whereas cognitive interaction diversity, whose differences approximated a normal distribution, was analyzed using a paired-sample t test.

**Analysis of Problem-Solving Pattern (RQ2).** To discover problem-solving patterns in which students engaged with the chatbot for problem solving, we used HINA's dyadic-level analysis [34] to identify the statistically significant interactions within the constructed graph in each condition. In this analysis, we excluded students' off-topic interactions to focus on the analysis of the statistically significant problem-solving patterns in the Human-AI interaction processes. We extracted a subgraph $G''$ from $G$. Specifically, $G''$ is also a bipartite network structure by removing *Irrelevant Conversations* nodes and their associated heterogeneous interactions from $G$. We fixed the degree of student node set $V_s$ and then pruned the $G''$ (set $\alpha = 0.01$), controlling for differences in student interaction intensity. This process ensures that the resulting network demonstrates statistically significant interactions relative to each student's overall engagement, preventing highly active students from dominating the network structure. The pruned networks in the two chatbot conditions were then compared to reveal distinct problem-solving patterns associated with the custom and general-purpose chatbots. The pruning process and visualization were conducted via the HINA Web Tool [35].

**Analysis of Problem-Solving Performance (RQ3).** Students' problem-solving performance was operationalized as solution scores. The solutions were evaluated by researchers for correctness and completeness, receiving scores on a 0-6 scale. All grades were reviewed and confirmed by a second grader independently. Descriptive statistics were first conducted to provide an overview of students' problem-solving performance across each task and chatbot condition. To examine the main effect of chatbot type while accounting for potential differences in task difficulty and order, a linear mixed-effects model was employed. In this model, chatbot type, task topic, and order were specified as fixed effects, and students were treated as a random effect. All statistical analyses were conducted using IBM SPSS Statistics.

# 4 Results

## 4.1 RQ1: Comparison of Interaction Intensity and Cognitive Interaction Diversity

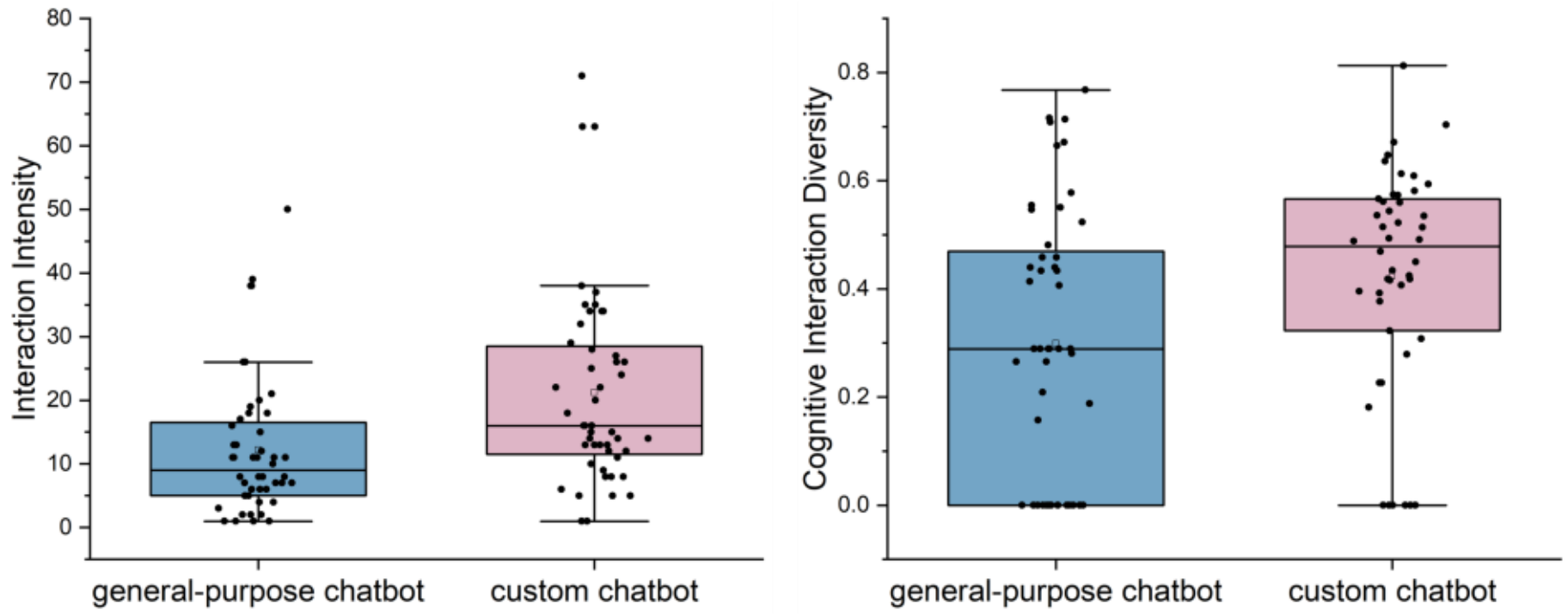

**Fig. 3.** Comparison of interaction intensity and cognitive interaction diversity

Based on the constructed HIN and Eq. (1) and Eq. (2), we calculated students' interaction intensity and cognitive interaction diversity. A Wilcoxon signed-rank test was conducted and we found a significant difference in students' interaction intensity between the two chatbot conditions ($Z = 4.087$, $p < 0.001$, $r = 0.590$), with students interacting more with the custom chatbot ($M = 21.17$) than with the general-purpose chatbot ($M = 12.21$). The results of a paired-samples t-test indicated a significant difference in students' cognitive interaction diversity ($t = 3.301$, $p = 0.004$, $d = 0.44$), with students exhibiting significantly greater cognitive interaction diversity with the custom chatbot ($M = 0.420$) compared to the general-purpose chatbot ($M = 0.299$), suggesting that students were more effectively guided by the custom chatbot to employ more diverse cognitive strategies. Fig. 3 visualized the distribution of interaction differences between the two groups of students through a boxplot.

## 4.2 RQ2: Comparison of Problem-Solving Pattern

Fig. 4(a) indicated that when using the custom chatbot, Follow steps was the dominant interaction, with nearly all students learning concepts and solving problems under the chatbot's guidance. Some students also *Refined solutions* based on the custom chatbot's feedback through reflection, whereas no students primarily engaged in refining solution with the general-purpose chatbot. Fig. 4(b) showed that with the general-purpose chatbot, *Direct Request* such as Copy & Paste and Request language translation were prominent across many students, and several students frequently requested the chatbot to format responses. Regarding socio-emotional interactions, students shared their current understanding and situational context more often when interacting with the custom chatbot. Overall, there were more significant *Exploratory Inquiry* interactions with the custom chatbot, reflecting that students were engaged in problem-solving with GAI chatbots, whereas the general-purpose chatbot showed more *Direct Request*.

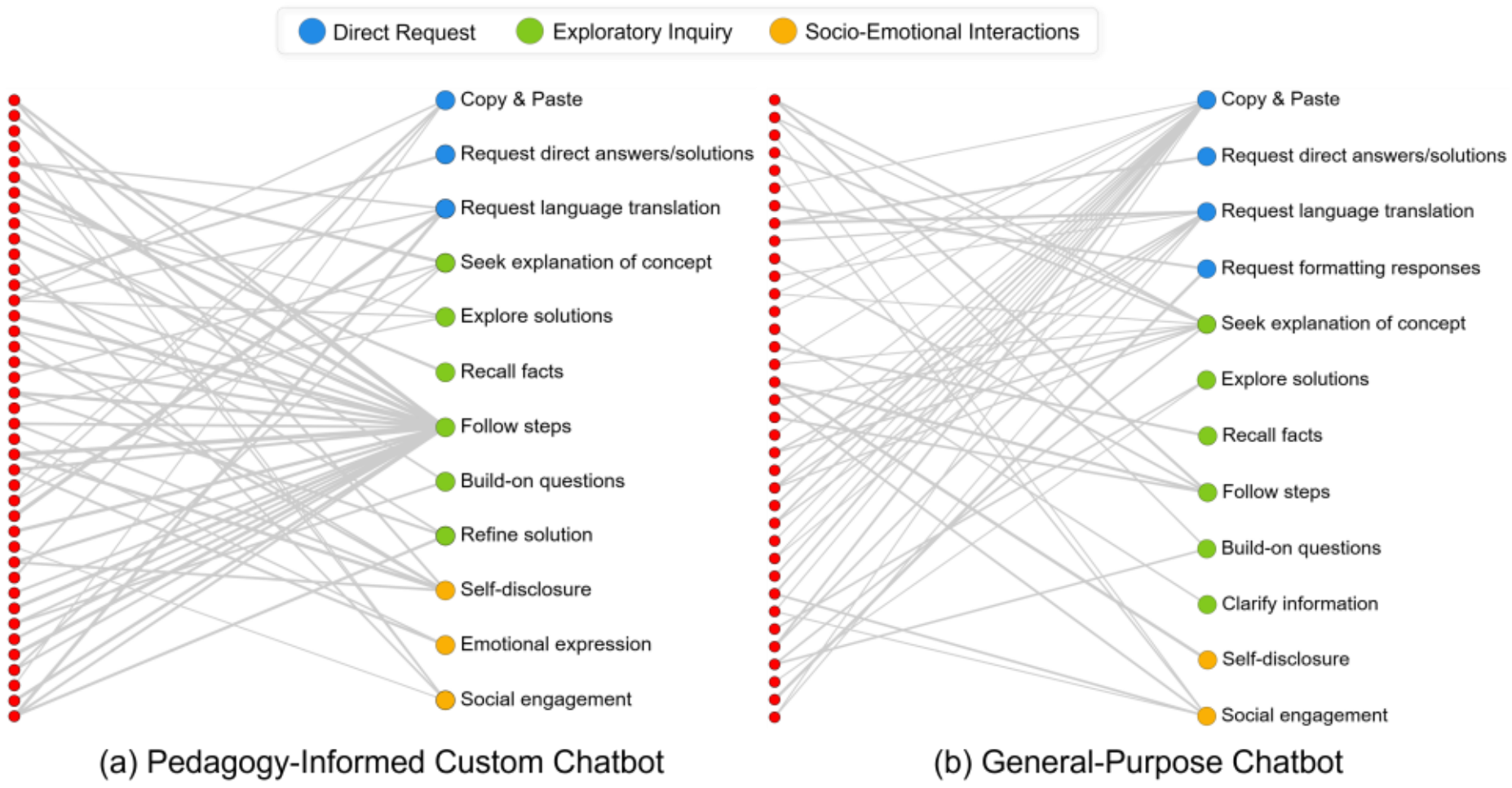

**Fig. 4.** Students' significant problem-solving patterns in science problem solving processes using the custom chatbot and the general-purpose chatbot. The two HINs in the graph, with nodes are colored by different interaction categories: red for students, blue for Direct Request, green for Exploratory Inquiry, and yellow for Socio-Emotional Interactions; edges between heterogeneous nodes indicate significant interactions during science problem solving.

### 4.3 RQ3: Comparison of Problem-Solving Performance

Descriptive statistics indicated that, across both tasks, students showed a slightly higher mean value of problem-solving performance with the custom chatbot (Task 1: $M = 3.30$, $SD = 1.74$; Task 2: M = 3.33, $SD = 1.57$) than with the general-purpose chatbot (Task 1: $M = 2.71$, $SD = 1.53$; Task 2: $M = 3.20$, $SD = 1.65$). Linear mixed model analysis was conducted to examine the effects of chatbot type on students' problem-solving performance. However, the statistical analysis results revealed there was no significant main effect for chatbot type ($F = 1.521$, $p = 0.224$). Furthermore, the main effect of task topic ($F = 0.825$, $p = 0.368$) and order ($F = 0.403$, $p = 0.529$) were also non-significance, suggesting that neither the specific task content nor the sequence of chatbot usage significantly biased the students' performance. The results of estimated marginal means (EMMs) derived from the model indicated that students performed better on average with the custom chatbot (EMM = 3.317) than with the general-purpose chatbot (EMM = 2.937) across tasks.

# 5 Discussion

## 5.1 Problem-Solving Interactions with Different GAI Chatbots

This study provides new theoretical insights and empirical evidence on how students interact differently with different types of GAI chatbots during problem-solving tasks, highlighting the importance of designing educational chatbots grounded in pedagogical theories and learning contexts. Although general-purpose chatbots are widely adopted due to their powerful generative capabilities, they may not always align with educational goals [8]. In this study, students frequently requested the general-purpose chatbot to execute specific commands—such as translating, giving answers, or formatting responses—outsourcing cognitive work that would otherwise require their own processing [36]. The substitution of cognitive work by GAI chatbots may reduce students' cognitive engagement, which may impede their sense-making [37]. Our findings echo prior research on cognitive offloading associated with GAI use, which has raised concerns about the potential risks of reduced cognitive effort and autonomy in human-AI interaction [37, 38].

In this study, we found that the custom chatbot, which provides Socratic questions instead of providing direct answers, was particularly effective in supporting students' meaningful problem-solving interactions. We found that students interacted more frequently with the custom chatbot and employed a wider range of cognitive strategies, reflecting a more active form of learning compared to the general-purpose chatbot. The Socratic prompts encouraged students to engage in multi-turn dialogues and gradually scaffolded them to adopt diverse cognitive strategies such as reasoning, summarizing, and explaining concepts, fostering deeper understanding and reflection [9, 40]. For science problem-solving tasks, which require intensive cognitive processing [1], the custom chatbot could provide stronger support. This observed difference reinforces the perspectives by Rivera-Novoa et al. [26], which suggests that in science learning, general-purpose chatbots often function as cognitive substitutes, whereas interactions with Socratic chatbots operate as a coupled and extended system, serving as a cognitive complement that positions students as active agents in their own learning processes. We also observed that the custom chatbot elicited more reflection and revision behaviors, encouraging students to regulate their own learning processes. Such metacognitive support can help mitigate metacognitive laziness, a phenomenon that students over-rely on GAI and offload metacognitive load [38]. Our findings underscore that pedagogy-informed customization of GAI is essential to mitigate its inherent risks and ensure alignment with pedagogical objectives.

## 5.2 The Impact of GAI Chatbots on Problem-Solving Performance

Although the custom chatbot significantly shaped students' HAI interactions in problem-solving processes, no significant difference was observed in problem-solving performance evaluated by the solution quality between the two chatbot conditions. Similar non-significant results were reported in prior studies [39]. This finding suggests a dissociation between problem-solving processes and outcome-level performance in a GAI-supported problem-solving context. In our context, the general-purpose chatbot is capable of producing largely complete solutions, offering a cognitive shortcut for students [7]. As a result, students may achieve comparable solution scores through copying the general-purpose chatbot's answers, even when their underlying cognitive interactions and science problem-solving ability differ. Similar concerns regarding the misalignment between AI-assisted performance and students' actual ability and knowledge level have been raised in prior work [38]. This finding underscores the necessity of examining students' problem-solving processes in a GAI-supported learning context, rather than focusing solely on outcomes.

### 5.3 Limitations and Future Work

While this study provides valuable insights, several limitations should be acknowledged. The participating school was a single-gender school for girls. Future studies are recommended to include mixed-gender samples to investigate the possibility of gender influence on GAI effects. Furthermore, this study operationalized problem-solving performance in terms of solution quality. There is a need for future research to examine students' knowledge gains and long-term knowledge retention.

## 6 Conclusion

In conclusion, our study found that pedagogy-informed custom GAI chatbots significantly fostered stronger interactions and more diverse cognitive interactions in science problem solving than the general-purpose chatbot. However, no significant differences were observed in problem-solving performance between two chatbot conditions. These findings demonstrated that the custom chatbot was effective in supporting students' science problem solving, while the general-purpose chatbot led to cognitive offloading. This study contributes to the field of GAI-supported science education and Human-AI Interaction, and motivates further research to explore best practices for integrating GAI to promote more effective and engaging student-AI interactions in science learning.